\documentclass[aps,prl,twocolumn,showpacs, fleqn]{revtex4}%superscriptaddress,

\usepackage{graphicx}% Include figure files
\usepackage{dcolumn}% Align table columns on decimal point
\usepackage{bm}% bold math
\usepackage{psfrag}% symbal display

\begin{document}

\title{Preparation of electron-doped La$_{2-x}$Ce$_{x}$CuO$_{4\pm \delta}$
thin films with various Ce doping by dc magnetron sputtering}

\author{B. X. Wu$^{1}$, K. Jin$^{1}$, J. Yuan$^{2}$, H.B. Wang$^{2}$,
T. Hatano$^{2}$, B. R. Zhao$^{1}$ and B. Y. Zhu$^{1}$%\corref{cor1}
}
%\ead{beiyi.zhu@aphy.iphy.ac.cn}

\affiliation{$^1$National Laboratory for Superconductivity,
Institute of Physics, and Beijing National Laboratory for Condensed
Matter Physics, Chinese Academy of Sciences, Beijing 100190, China}

\affiliation{$^2$National Institute for Materials Science, Tsukuba
305-0047, Japan}

\begin{abstract}
A series of $c$-axis oriented electron-doped high-T$_c$
superconducting La$_{2-x}$Ce$_{x}$CuO$_{4}$ thin films, from heavily
underdoped $x$=0.06 to heavily overdoped $x$=0.19, have been
synthesized by dc magnetron sputtering technique on $(100)$
SrTiO$_{3}$ substrates. The influence of various fabrication
conditions, such as the deposition temperature and the deposition
rate, etc., on the quality of the thin films has been scrutinized.
We find that the quality of the films is less sensitive to the
deposition temperature in the overdoped region than that in the
underdoped region. In the phase diagram of T$_c(x)$, the
superconducting dome indicates that the optimally doping level is at
the point $x=0.105$ with the transition temperature T$_{c0} =
26.5$~K. Further more, both the disappearance of the upturn in the
$\rho_{xx}$(T) curve at low temperature under H=10~T and the
positive differential Hall coefficient, $R_H'=d \rho_{xy}/dH$, are
observed around $x = 0.15$, implying a possible rearrangement of
Fermi surface at this doping level.
\end{abstract}


\maketitle

\pagebreak

\section{1. Introduction}
Electron-doped cuprate superconductors have attracted more and more
attention since its discovery~\cite{Tokura}. This type of cuprates
may have a great impact on our understanding of the mechanism of
high temperature superconductivity, not only because of the common
properties shared with the hole-doped cuprates, but also due to its
unique features. The common points are that (i) the parent compounds
are perovskite Mott insulators; (ii) the CuO$_2$ plane is the key
element responsible for the superconductivity~\cite{Orenstein};
(iii) two types of carriers coexist~\cite{Leboeuf, Greene, kuiJ1}.
The above items imply the electron-hole symmetry in the cuprate
superconductors. However, the unique properties of the
electron-doped cuprates are as following: (i) broad
antiferromagnetic phase doping range, which can extend to or even
coexist with the superconducting (SC) phase in some sense, (ii)
narrow SC dome~\cite{Devereaux}, (iii) Fermi liquid behavior with a
quadratic rather than linear temperature dependence of the
resistivity~\cite{Zhao,WH}, (iv) the order parameter symmetry, which
does not come to a general consensus on electron-doped cuprates,
while it has been accepted as d-wave type on the hole-doped
ones~\cite{Tsuei, Sato, Armitage1}, (v) low SC transition
temperature T$_c$ and low upper critical field H$_{c2}$, which give
us the opportunity to suppress the superconductivity even at
extremely low temperature by the magnetic field and explore the
normal state properties, e.g., the quantum phase transition and the
anomalous upturn of the low-temperature resistance~\cite{Li,
Dagan1}.

Ln$_{2-x}$Ce$_x$CuO$_4$ (Ln=Nd, Pr, Sm, and Eu) family with the
so-called T'-phase structure is the extensively explored material in
the electron-doped cuprates. Among them, La$_{2-x}$Ce$_x$CuO$_4$
(LCCO), with the largest Ln$^{3+}$ ion radius~\cite{Naito1, Cooper,
Arima}, has the highest transition temperature up to about 30~K.
However, due to the difficulty to get rid of the excess oxygen to
achieve the superconduting T' phase~\cite{phase}, LCCO was first
synthesized by Yamada~\cite{Yamada} in 1994 by a rather complicated
precursor technique. Recently, Naito {\it et al.} synthesized
superconducting T'-LCCO films by molecular beam epitaxy
(MBE)~\cite{Naito2}. Sawa {\it et al.} got various doped LCCO thin
films on SrTiO$_3$ substrate using BaTiO$_3$ as a buffer layer by
the pulsed laser deposition (PLD) method~\cite{Sawa}. Zhao {\it et
al.} also successfully grew the optimal doped LCCO thin film on
SrTiO$_{3}$ by dc magnetron sputtering~\cite{Zhao}. In this paper,
we systematically study the fabrication conditions for the c-axis
oriented LCCO thin films on SrTiO$_3$ substrates by dc magnetron
sputtering method, including the deposition temperature and the
deposition rate for various Ce doping levels, ranging from heavily
underdoped to heavily overdoped, i.e. $x = 0.06 \sim 0.19$. Besides
that, the optimal deposition conditions are discussed in detail. The
SC dome in the phase diagram T$_c(x)$ is obtained and the optimal
doping level is at $x$=0.105 with the highest transition temperature
T$_{c0}$=26.5~K. By investigating the evolution of the resistivity
$\rho_{xx}$ and the differential Hall coefficient R$_H'$ with the
doping level $x$, we find that the rearrangement of the Fermi
surface (FS) at the doping level $x$=0.15.

\section{2. Experiments}
We prepared a series of $c$-axis thin film
La$_{2-x}$Ce$_{x}$CuO$_{4\pm \delta}$ (LCCO) with different Ce
doping concentration $x$ ranging from 0.06 to 0.19 on the (100)
oriented SrTiO$_{3}$ by dc magnetron sputtering method. The targets
were synthesized by conventional solid-state reaction~\cite{Zhao}.
To obtain the high quality LCCO thin films, we carefully adjusted
the sputtering pressure, the deposition temperature, the gas ratio
of Ar to O$_2$, the deposition rate and the annealing process for
each doping. The background pressure of the chamber prior to
deposition was less than $3.0 \times 10^{-4}$~Pa. The total pressure
during the deposition was $40-50$~Pa with O$_{2}$:Ar=1:4. The
deposition temperature T$_D$ for the LCCO thin film is in the range
of $620 \sim 750^{\circ}$C, depending on the Ce doping level. More
specific studies focusing on the optimal deposition temperature at
different doping levels are investigated in part 3. All the films
used in the present work are typically about 100 nm in thickness,
and they were patterned into the standard micro-bridge with 1~mm
long and 500~$\mu$m wide by photolithography and ion milling
techniques. All the transport measurements were carried out by the
Quantum Design PPMS-14 equipment.

\section{3. Results and discussions}
\begin {figure}[!]  %\graph1
\begin{center}
\includegraphics*[bb=5 141 450 665, width=8cm, clip]{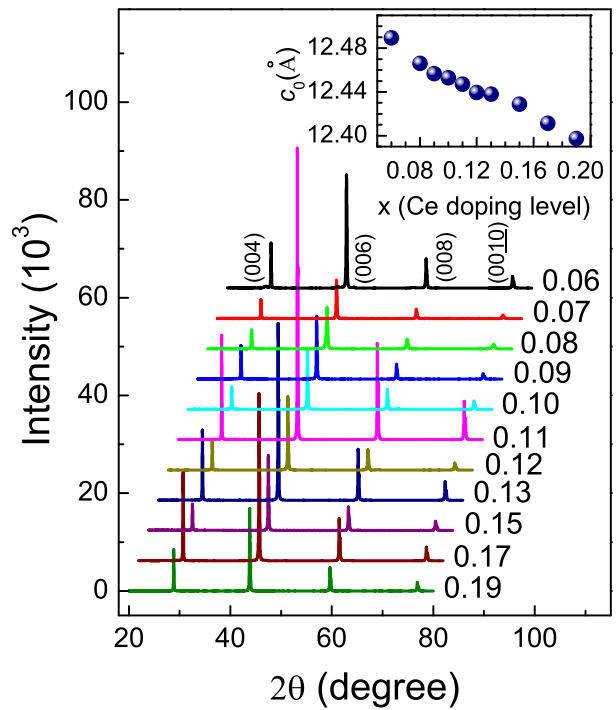}
\caption{\label{XRD} X-ray diffraction diagram for the LCCO films
with different Ce doping levels $x$ from 0.06 to 0.19. The substrate
peaks are removed. The inset shows the dependence of the $c$-axis
lattice parameter $c_{0}$ on the Ce doping level $x$.}
\end{center}
\end{figure}

Fig.~\ref{XRD} shows the X-ray diffraction (XRD) data of the LCCO
thin films with $x=0.06 \sim 0.19$. All the samples are synthesized
at their optimal deposition temperatures for the highest T$_c$,
respectively. As shown in Fig.~\ref{XRD}, all the LCCO thin films
are in $c$-axis orientation with $(00l)$ peaks signed. The results
agree with those obtained by other methods, such as MBE and
PLD~\cite{Naito2,Sawa}. In detail, the remain of the T-phase in the
LCCO samples can be disclosed by the tiny humps near the sharp peak
of T'-(004) and (006) in the heavily underdoped samples at x=0.06.
The inset of Fig.~\ref{XRD} presents the dependence of the $c$-axis
lattice parameter $c_{0}$ on the doping level $x$. $c_{0}$ decreases
monotonically from 12.48~\AA~to 12.39~\AA~with $x$ increasing. This
can be attributed to the fact that the atom radius of Ce is smaller
than that of La.

\begin {figure}[!]  %\graph2
\begin{center}
\includegraphics*[bb=29 290 394 630, width=8cm, clip]{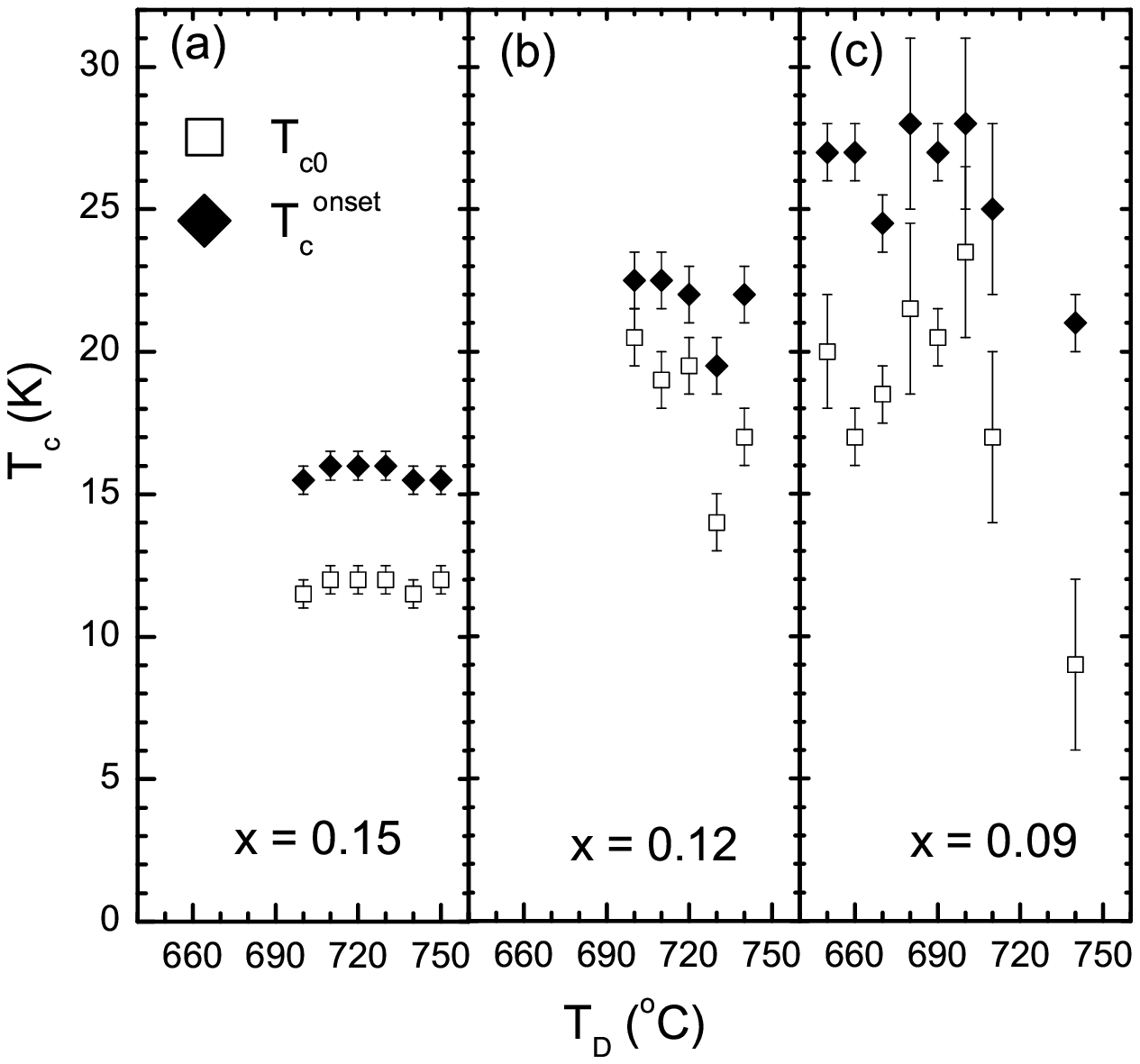}
\includegraphics*[bb=16 311 419 604, width=8cm, clip]{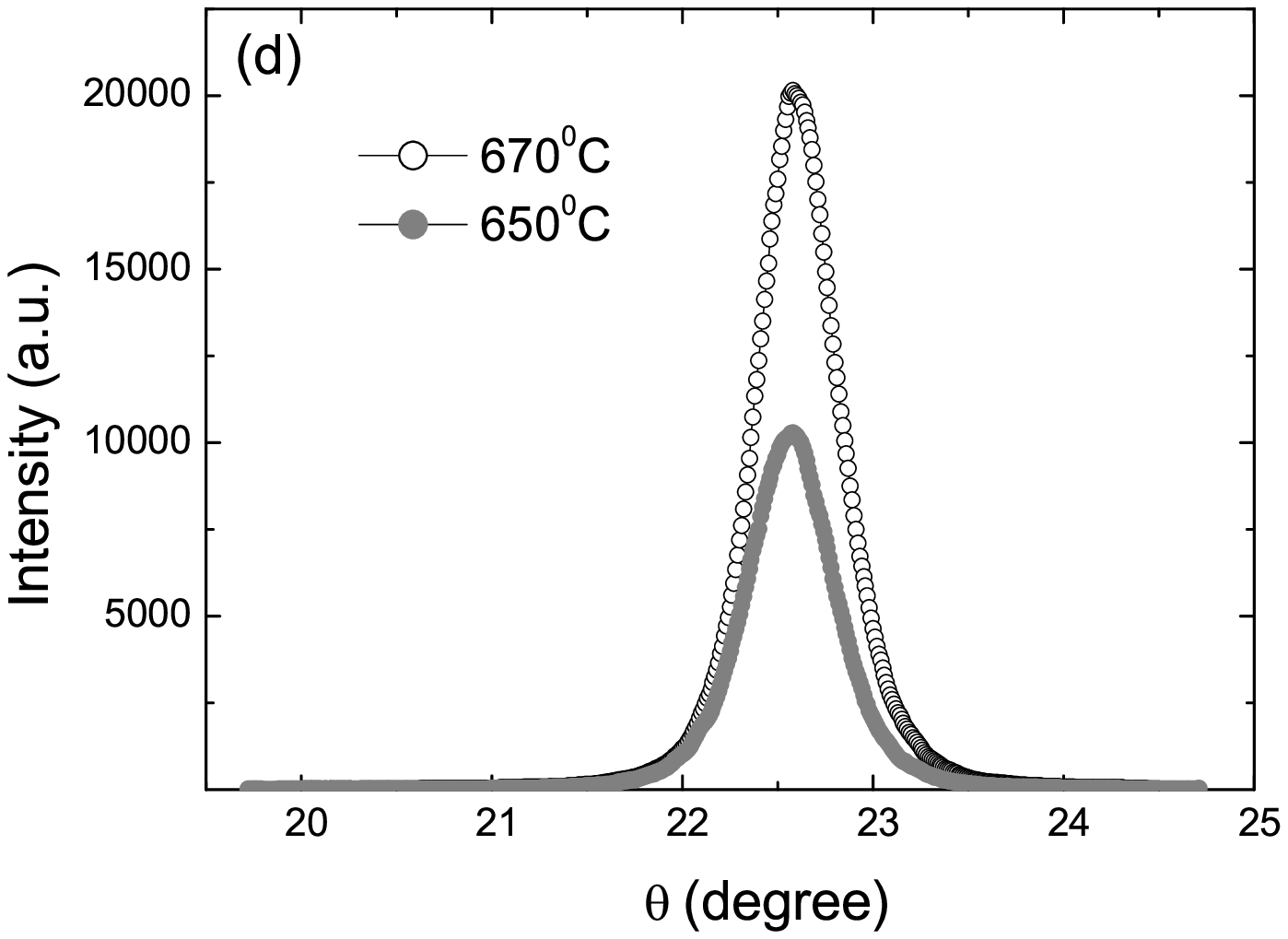}
\caption{\label{Tc} $T_{c0}$ and $T_c^{onset}$ of LCCO thin film
with three different doping levels, (a) $x=0.15$, (b) $x=0.12$, and
(c) $x=0.09$, deposited at various temperatures. The error bar is
determined by the variance of the reproducibility, and the
uncertainty of the T$_c$ due to the unsharp transition is also
included as considering the error bar at underdoped region. (d) The
rocking curves of the $(006)$ peaks for the films with $x=0.08$
fabricated at different deposition temperatures T$_D$.}
\end{center}
\end{figure}

Fig.~\ref{Tc}(a)-(c) shows both the onset and the zero resistance
superconducting transition temperature, T$_c^{onset}$ and T$_{c0}$,
for the LCCO films deposited at various temperatures with three
different doping levels, i.e. $x$=0.15, 0.12 and 0.09, respectively.
The error bar is determined by the variance of the reproducibility,
and the uncertainty of the T$_c$ due to the slight upturn of the
$\rho$(T) curve is also included as concerning the error bar at
underdoped region. For the case of $x$=0.15, both T$_{c0}$ and
T$_c^{onset}$ of the films deposited around $710 \sim 730^{\circ}$C
have the maximum values, i.e. 12~K and 16~K, respectively. When T$_D
= 700^{\circ}$C, 740$^{\circ}$C and 750$^{\circ}$C, both T$_c$ and
T$_c^{onset}$ decrease within $\sim 0.5$~K. Since this variance is
rather small, the optimal T$_D$ for $x$=0.15 is in a quite wide
temperature region $700 \sim 750^{\circ}$C with good
reproducibility. For $x = 0.12$, the best deposition temperature is
between 700$^{\circ}$C and 720$^{\circ}$C, where T$_c$ shows small
variation within $\sim 1$~K. While, if the film deposited at higher
temperature, T$_c$ decreases obviously. The films with $x$=0.09 are
fabricated at T$_D =650 \sim 740^{\circ}$C, and the transition
temperature T$_c$ shows stronger dependence on the deposition
temperature T$_D$ with bad reproducibility as seen in
Fig.~\ref{Tc}(c). The films have both the highest T$_c$ and the
sharpest transition width when deposited around the optimal T$_D =
680\sim700^{\circ}$C. The structure information shown in
Fig.~\ref{Tc}(d) gives the rocking curve of the (006) peak for the
films of $x$=0.08 deposited at 650$^{\circ}$C and 670$^{\circ}$C,
respectively. The full-widths at the half maximum (FWHM) of (006)
peak are about 0.42 for the sample deposited at 650$^{\circ}$C and
0.37 at 670$^{\circ}$C, which indicate that the quality of the film
deposited at 670$^{\circ}$C is better than the one deposited at
650$^{\circ}$C.

Both the transport and structure data show that the quality of the
underdoped films is affected strongly by the deposition temperature
T$_D$. And the high quality films can only be achieved in a small
deposition temperature region. The quality of the underdoped samples
is much more sensitive to T$_D$ than that of the overdoped ones, and
the reproducibility becomes worse for the samples with lower doping.
The XRD data in Fig.~\ref{XRD} also show the unexpected T-phase in
the heavily underdoped sample, e.g. $x$=0.06, which is hard to
eliminate no matter how we adjust the deposition condition, and it
agrees with the fact that there are strict restrictions for the
deposition in the underdoped region~\cite{Naito2,Sawa}.

Moreover, we find that the optimal deposition temperature for the
films decreases with the doping level decreasing from the overdoped
to the underdoped region as shown in Fig.~\ref{Tc}(a)-(c). For the
LCCO material, the analysis of the perovskite crystallographic
Goldschmidt tolerance factor $t$ indicates that the
non-superconducting T-phase tends to be built because the SC
T'-phase is unstable at relatively high synthesis
temperatures~\cite{phase}. Manthiram and Goodenough~\cite{tolerant}
predicted that T'-phase can only be stabilized below 425$^{\circ}$C,
while partial substitution of La$^{3+}$ by smaller Ce$^{4+}$ can
reduce $t$ and shift the T/T'-phase boundary to a higher temperature
about 600$^{\circ}$C for $x$=0.15. However, 600$^{\circ}$C is still
too low to prepare the bulk material. This can be the reason for the
fact that the optimal deposition temperature tends to be higher with
the increase of the Ce doping level $x$.

\begin {figure}[!]  %\graph3
\begin{center}
\includegraphics*[width=8cm, clip]{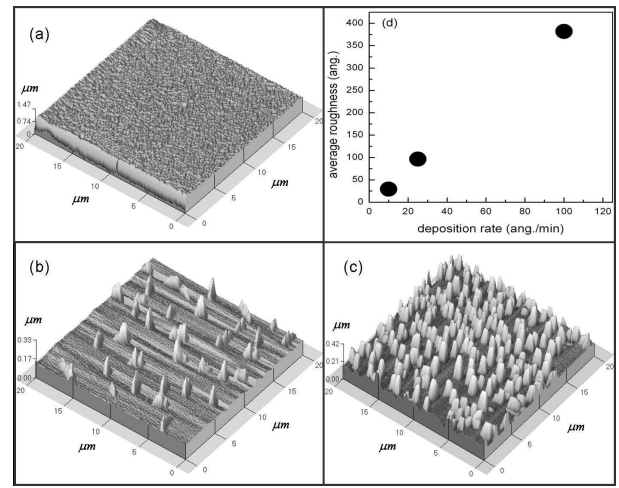}
\caption{\label{AFM}3D AFM  pictures of LCCO thin films ($x$=0.105)
deposited at different deposition rates: (a) 10~\AA/min, (b)
25~\AA/min, and (c) 100~\AA/min. The scanning region for each
picture is 20~$\mu$m$\times$20~$\mu$m. All the films are synthesized
at their optimally deposition temperature T$_D\sim700^{\circ}$C with
the same thickness of about 100~nm. (d) The evolution of the
roughness with the deposition rate.}
\end{center}
\end{figure}

Fig.~\ref{AFM} shows the atomic force microscope (AFM) 3D images of
the surface for the LCCO thin films $x$=0.105 synthesized at their
optimally deposition temperature T$_D\sim700^{\circ}$C with
different deposition rates. Each scanning region is $20 \mu$m$\times
20 \mu$m. The deposition rates for the samples in
Fig.~\ref{AFM}(a)$\sim$(c) are 10~\AA/min, 25~\AA/min and
100~\AA/min, respectively. We tune the deposition time for each rate
to assure all the films with the same thickness of about 100~nm. The
deposition time for films in Fig.~\ref{AFM}(a) to (c) is chosen as
100~min, 40~min and 10~min, respectively. We can see that the
surface of film in Fig.~\ref{AFM}(a) is quite smooth with only a few
grains scattered in the area. With the increase of the deposition
rate, a large quantity of ions will be accelerated and rush to the
substrate with a large velocity, which lead to an extended glow
plasma during the sputtering. Therefore, most of the arrived ions
may not have enough relaxation time to build the desirable epitaxial
structure in the film surface, which will result in the imperfect
features in the samples as shown in Fig.~\ref{AFM}(b) and (c).
Furthermore, the surface of film in Fig.~\ref{AFM}(c) is covered
with many big grains and the film almost grows in island rather than
layer-by-layer at this quite large deposition rate 100~\AA/min.
Fig.~\ref{AFM}(d) shows that the roughness of the film increases
almost monotonically with the deposition rate. Importantly, the
films deposited at about 25~\AA/min have the highest T$_c$, and any
deflection of the deposition rate will result in lower T$_c$. This
is helpful for us to fabricate the desirable films with either the
smooth surface or the highest T$_c$ with a tolerable roughness.

\begin {figure}[!]  %\graph4
\begin{center}
\includegraphics*[bb=55 401 461 742, width=8cm, clip]{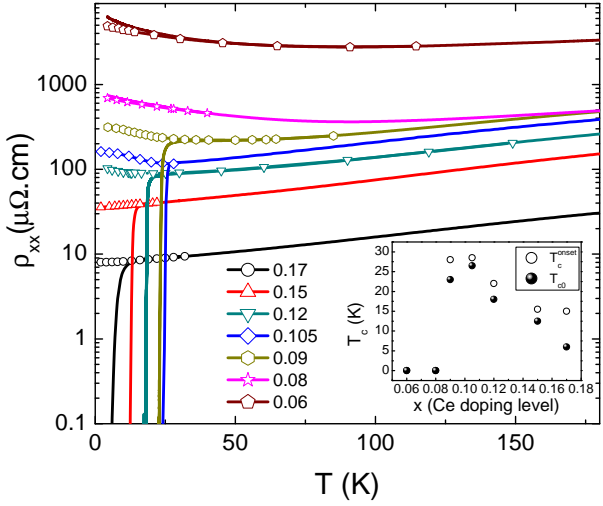}
\caption{\label{RT} The {\it ab}-plane resistivity dependence on the
temperature for the films with different doping levels $x$. The
solid lines and the open symbols represent the measurements at H=0~T
and H=10~T (H $>$ H$_{c2}$), respectively. Inset: phase diagram of
the transition temperature T$_c$ versus the doping level $x$.}
\end{center}
\end{figure}

A series of $c$-axis oriented thin films of LCCO with various Ce
doping levels, from heavily underdoped to heavily overdoped, were
deposited at the same deposition rate of 25~\AA/min with 40 minutes,
which results in the same thickness of about 100 nm for each film.
The resistivity versus the temperature curves for the various doping
levels have been measured as seen in Fig.~\ref{RT}. The
$\rho_{xx}$(T) curves show that the optimal doping level is at
$x$=0.105 with the highest transition temperature T$_{c0}$=26.5~K
and the sharpest transition width $\triangle$T $\leq$ 1.5~K. The
phase diagram is shown in the inset of Fig.~\ref{RT}. The results
are consistent with those reported by Naito {\it et
al.}~\cite{Naito2}. The optimal doping concentration in LCCO is
lower than those of Pr$_{2-x}$Ce$_x$CuO$_4$ and
Nd$_{2-x}$Ce$_x$CuO$_4$, where $x$=0.15 for both materials. It is
clear that the residual resistivity of the samples decreases with
the increase of $x$. The dependence of the normal state resistivity
on temperature shows an insulator-metal transition near the
optimally doped concentration $x$=0.105. As seen in the phase
diagram T$_c(x)$ in the inset of Fig.~\ref{RT}, there is a sharp
drop of T$_c$ when the doping level is biased from the optimal
doping level $x$=0.105 to the underdoped regime, and the
superconductivity is suppressed at $x$ $<$ 0.08. While in the
overdoped region, it decreases more slowly and the superconductivity
disappears at $x>0.19$.

\begin {figure}[!]  %\graph5
\begin{center}
\includegraphics*[bb=14 324 416 670, width=6cm, clip]{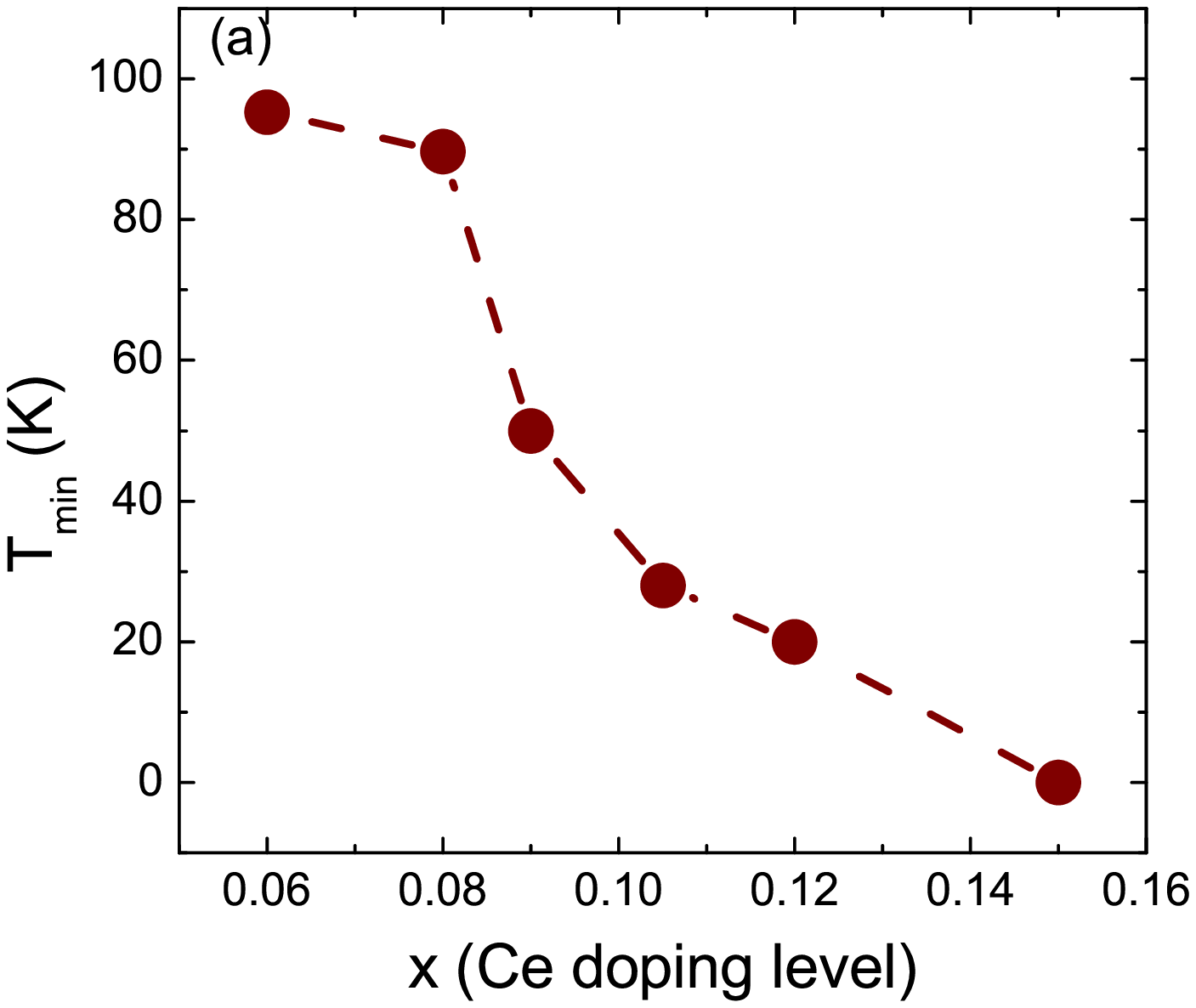}
\includegraphics*[bb=24 388 506 789, width=6cm, clip]{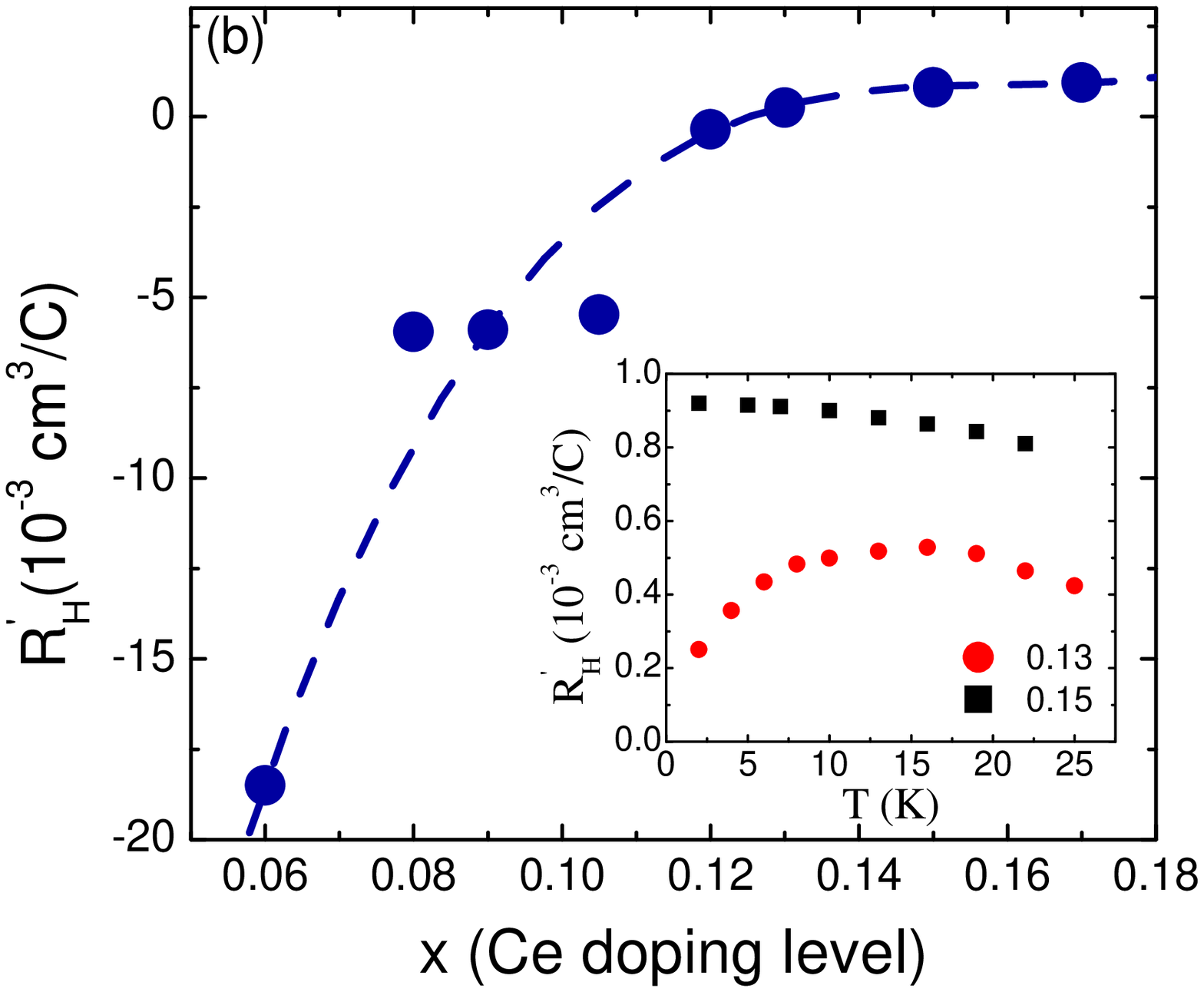}
\caption{\label{PT}(a) Temperature for the minimum resistivity and
(b) differential Hall coefficient $R_H'=d \rho_{xy}/dH$ at T = 2~K,
versus the doping concentration $x$. The inset of (b) shows the
temperature dependence of the differential Hall coefficient R$_H'$
for $x$=0.13 (red circle) and 0.15 (black square). The dashed lines
are the guides to the eyes.}
\end{center}
\end{figure}

The $\rho_{xx}$(T) curves under H=10~T, as shown with the open
symbols in Fig.~\ref{RT}, exhibit anomalous upturn at low
temperature, which is related to the carrier
localization~\cite{Dagan1,WL1,Kondo1,Kondo2}. When the
superconductivity is suppressed by an applied field, e.g. H=10~T,
the temperature T$_{min}$, where the sample has the minimum
resistivity, decreases with the increase of the doping concentration
$x$, and it becomes zero in the overdoped side at $x$=0.15, as seen
in Fig.~\ref{PT}(a). Moreover, we have studied the low temperature
Hall effect of the LCCO thin films at various doping levels, which
could reflect the electronic structure of the material in some sense
due to its free from complicated inelastic
scattering~\cite{Kui1,Kui2}. In Fig.~\ref{PT}(b), we show the Ce
doping dependence of the differential Hall coefficient R$_H'$ at
2~K, obtained from the slope of a linear fit to the Hall data at
H=10~T, i.e. $R_H'=d \rho_{xy}/dH$. R$_H'$ increases from a strong
negative value to a positive value with the increase of $x$, and a
minor positive R$_H'$ starts to emerge at $x\sim$0.13. However, the
temperature dependence of R$_H'$ for the films with $x$=0.13 shows a
tendency to be negative when the temperature decreasing lower than
16~K, as seen in the inset of Fig.~\ref{PT}(b). This indicates that
the coexistence of the hole-like and electron-like Fermi pockets,
which may be formed by the intersection of the FS and the
antiferromagnetic Brillouin Zone~\cite{Armitage2}, is still resident
in this doping level at low temperature. On the other hand, for
$x$=0.15, R$_H'$ is positive in the whole temperature region from
2~K to 300~K, and saturates to a positive value when T decreasing
close to zero, which implies the formation of the large hole-like FS
at this doping. Therefore, both the T$_{min}$ and the R$_H'$ data
indicate that the rearrangement of the FS takes place at the doping
level $x$=0.15, compatible to the results obtained in
Pr$_{2-x}$Ce$_x$CuO$_4$~\cite{Dagan}. We noticed that there were
similar results in both the electron-doped cuprates and the
hole-doped cuprates. An antiferromagnetic phase starts at $x$=0 and
extends to the SC dome in the electron-doped cuprates~\cite{Luke}. A
low temperature phase transition from insulator to metal as a
function of doping is observed in hole-doped
materials~\cite{Boebinger} and the evolution of the FS versus the
doping concentration has also been revealed by the ARPES
method~\cite{Armitage2,Matsui}. More efforts on the FS evolution
with doping concentration at extremely low temperature are still
necessary for the further understanding of the cuprates
superconductors.

\section {4. Conclusion}
We have prepared a series of LCCO thin films with various Ce doping
levels ($x$=0.06$\sim$0.19). The XRD results indicate that the LCCO
thin films are of high quality. The influence of the deposition
conditions is discussed in detail. Compared with the overdoped
films, the underdoped films are hard to be synthesized due to the
crucial restriction on the deposition temperature, which can be
attributed to the instability caused by the emergence of the T-phase
in the heavily underdoped region. Furthermore, the optimal
deposition temperature increases with the increase of doping
concentration $x$. We also find that a large deposition rate may
lead to a rough surface due to the lack of enough relaxation. The
deposition rate at around 25~\AA/min can achieve a high-T$_c$ film
with acceptable roughness, and either small or large deposition rate
will result in a lower-T$_c$ value. The SC dome in the phase diagram
indicates the optimal doping at $x$=0.105 with T$_{c0}$=26.5~K and a
narrow transition width $\triangle$T=1.5~K. Both the longitudinal
resistivity and differential Hall coefficient R$_H'$ versus the
doping level are discussed, and a possible rearrangement of the FS
is revealed at $x$=0.15.

\section{Acknowledgments}

We thank B. Xu, S.L. Jia, W.W. Huang and H. Chen for their help with
the measurements. The work is supported by the MOST, NSF, SRF for
ROCS, SEM Projects of China.

\end{document}